\documentclass{aa}  
\usepackage{graphicx,caption}
\usepackage{txfonts}

\usepackage{xcolor}
\usepackage{threeparttable}
\usepackage[colorlinks=true, allcolors=blue]{hyperref}

\usepackage{natbib}
\bibpunct{(}{)}{;}{a}{}{,}

\newcommand{\ergcm}[1]{$\times 10^{#1}$ erg cm$^{-2}$ s$^{-1}$}
\newcommand{\oergcm}[1]{$10^{#1}$ erg cm$^{-2}$ s$^{-1}$}
\newcommand{\ergs}[1]{$\times 10^{#1}$ erg s$^{-1}$}
\newcommand{\oergs}[1]{$10^{#1}$ erg s$^{-1}$}
\newcommand{\hcm}[1]{$\times 10^{#1}$ cm$^{-2}$}

\newcommand{\xspec}{\texttt{XSPEC}\xspace}

\newcommand{\swift}{{\it Swift}\xspace}

\newcommand{\nus}{{\it NuSTAR}\xspace}
\newcommand{\nustar}{{\it NuSTAR}\xspace}

\newcommand{\cxo}{{\it Chandra}\xspace}
\newcommand{\fermi}{{\it Fermi}\xspace}

\newcommand{\nicer}{{\it NICER}\xspace}

\newcommand{\sxp}{SXP\,15.6\xspace}

\graphicspath{{./}{plots_F/}}

\begin{document} 

\title{An X-ray view of the 2021 outburst of SXP 15.6: constraints on the binary orbit and magnetic field of the Neutron Star}

\author{G. Vasilopoulos\inst{\ref{oas}} \and
        G.~K.~Jaisawal\inst{\ref{nsid}} \and
        C. Maitra\inst{\ref{mpe}} \and
        F. Haberl\inst{\ref{mpe}} \and
        P. Maggi\inst{\ref{oas}} \and
        A.~S.~Karaferias\inst{\ref{uoa}}
       } 

\titlerunning{2021 outburst of SXP 15.6}
\authorrunning{Vasilopoulos et al.}

\institute{
Universit\'e de Strasbourg, CNRS, Observatoire astronomique de Strasbourg, UMR 7550, F-67000 Strasbourg, France\label{oas}\\ \email{georgios.vasilopoulos@astro.unistra.fr},
\and
National Space Institute, Technical University of Denmark, Elektrovej 327-328, DK-2800 Lyngby, Denmark\label{nsid},
\and
Max-Planck-Institut f{\"u}r extraterrestrische Physik, Gie{\ss}enbachstra{\ss}e 1, 85748 Garching, Germany\label{mpe}, 
\and
Department of Physics, National and Kapodistrian University of Athens, University Campus Zografos, GR 15783, Athens, Greece\label{uoa}} 
% \and
% Space Science Division, U.S. Naval Research Laboratory, Washington, DC 20375, USA \label{nrl}}

\date{Received ... / Accepted ...}

\abstract
   {} % leave it empty if necessary  
   {We conducted a spectral and temporal analysis of X-ray data from the Be X-ray binary pulsar SXP 15.6 located in the Small Magellanic Cloud based on \nustar, \nicer and \swift observations during the 2021 outburst.}
   {We present for the first time the broadband X-ray spectra of the system based on simultaneous \nustar and \nicer observations. Moreover we use monitoring data to study the spectral and temporal properties of the system during the outburst.
   }
   {Comparison of the evolution of the 2021 outburst with archival data reveals a consistent pattern of variability with multiple peaks occurring at time intervals similar to the orbital period of the system ($\sim$36 d).
   Our spectral analysis indicates that most of the energy is released at high energies above 10 keV while we found no cyclotron absorption line in the spectrum. Analysis of the spectral evolution during the outburst, we find that the spectrum is softer-when-brighter, which in turn reveals that the system is probably in the super-critical regime where the accretion column is formed. This places an upper limit to the magnetic field of the system of the order of 7$\times$10$^{11}$~G.
   The spin-evolution of the neutron star (NS) during the outburst is consistent with an NS with a low magnetic field ($\sim{5}\times$10$^{11}$~G), while there is evident orbital modulation which we modelled and derived the orbital parameters. We found the orbit to have a moderate eccentricity of $\sim$0.3.
   Our estimates of the magnetic field are consistent with the lack of an electron cyclotron resonance scattering feature in the broadband X-ray spectrum.
   }
   {}

\keywords{Galaxies: Magellanic Clouds --
          X-rays: binaries --
          Stars: neutron --
          X-rays: individual: \sxp
         }

\maketitle   

%________________________________________________________________

\section{Introduction}
\label{sec:intro}

Accreting X-ray pulsars (XRPs) in high-mass X-ray binaries (HMXBs) are of key importance in the study of accretion and binary evolution. 
The majority of XRPs are found in the so-called Be X-ray binaries \citep[BeXRBs; see][for a review]{2011Ap&SS.332....1R}, where mass transfer from the donor to the neutron star (NS) occurs through a slow moving equatorial disk (i.e. decretion disk).
A plethora of information about the physical properties of the systems may be acquired during outburst. Type I outbursts occur during the NS periastron passage, while major outbursts with X-ray luminosity $L_{X}>$\oergs{37} are quite rare and occur on timescales of years to decades \citep[][]{2013PASJ...65...41O,2014ApJ...790L..34M,2021ApJ...922L..37M}. These rare outbursts enable the study of the broad band X-ray spectral shape and the search for cyclotron resonant scattering features (CRSFs), which offer the only tool to directly measure the magnetic field at the NS surface. During outbursts, such systems have been found to show both evolution in the continuum spectra and --  in some cases -- detect CRSFs \citep{2016MNRAS.461L..97J,2018MNRAS.480L.136M,2019A&A...622A..61S}, yielding diagnostic data for the state of accretion \citep{2007ApJ...654..435B,2012A&A...544A.123B,2015MNRAS.452.1601P}.
Although the Magellanic Clouds (MCs), especially the Small Magellanic Cloud (SMC), are populous in BeXRBs \citep{2016A&A...586A..81H}, the presence of a CRSF is known only for a handful of systems, mainly owing to the lack of coverage in the hard X-ray band during its bright state. 
Moreover, X-ray pulsars in the MCs offer quite favourable conditions as they are placed at known distances (in contrast to most Galactic systems) and suffer from low Galactic absorption. This allows a precise determination of their luminosities.
In addition, the knowledge of the spin period and the magnetic field is crucial to constrain the accretion torque models and examine whether most of the BeXRBs are in spin equilibrium. 
Even if the magnetic field cannot be directly measured via CRSF, the study of the spin-evolution can deliver constraints or indirect estimates of the magnetic field of the NS.

\sxp (also known as XMMU\,J004855.5$-$734946) is a  BeXRB pulsar in the SMC \citep{2017MNRAS.470.4354V}. The orbital period of the system has been proposed to be 36.432 d based on optical monitoring data from OGLE \citep{2017MNRAS.467.1526M}, while a more recent analysis of OGLE data derived an optical period of 36.411 d \citep{2022MNRAS.513.5567C}.
The spin period of the NS was detected in 2016 from \cxo observations and since then no further strong outburst was witnessed until 2021. On 2021 November 19 a strong outburst was detected by \swift/XRT \citep{2021ATel15054....1C} at a luminosity level of 2\ergs{37} (0.3--10 keV) for a distance of 62 kpc \citep[][]{2014ApJ...780...59G}. Follow up \nicer and \nustar target of opportunity (ToO) observations enabled us to study the spin evolution of the NS as well as its broadband spectral properties. Based on those data we report on estimates of fundamental properties of the system like the orbital parameters and the magnetic field of the NS.

\section{The 2021 outburst of SXP 15.6}
\label{sec:outb}

The 2021 outburst was reported by \citet{2021ATel15054....1C} as a result of monitoring from the \swift Small Magellanic Cloud (SMC) Survey \citep[S-CUBED][]{2018ApJ...868...47K}. S-CUBED, is a high-cadence (1-2 weeks) shallow X-ray survey of the SMC that consists of $\sim$140 tiled pointings covering the optical extent of the SMC. The survey has been responsible for the early detection of some of the brightest outbursts in the MCs \citep[e.g. SMC X-3][]{2016ATel.9362....1K}. Following the announcement of the outburst, \nicer monitoring observations were performed, while \nustar observed the system near its peak flux with a ToO observation. In the following paragraphs we provide information for the \nicer and \nustar data that were collected during the outburst as well as \cxo archival data that were used for comparative studies.

\subsection{Data analysis}
\label{sec:data}

\subsubsection{NuSTAR}
The Nuclear Spectroscopic Telescope Array (\nustar) mission carries the first focusing high-energy X-ray telescope in orbit operating in the band from 3 to 79 keV \citep{2013ApJ...770..103H}.
\nustar observed the system with a 42~ks DDT observation (obsid: 90701339002) on 2019 November 26 (MJD 59544.40-59545.25).
\nustar data were analysed with version 1.8.0 of the \nustar data analysis software (DAS), and instrumental calibration files from {\tt CalDB} v20220301. The data were calibrated using the standard settings on the {\tt NUPIPELINE} script, reducing internal high-energy background, and screening for passages through  the South Atlantic Anomaly. We used the {\tt NUPRODUCTS} script to extract phase-averaged spectra for source and background regions (60\arcsec\ radius) for each of the two focal plane modules (FPMA/B). 
Finally, we performed barycentric corrections to the event times of arrival using the satellite's orbital ephemeris files.

\subsubsection{NICER}

The \nicer X-ray Timing Instrument \citep[XTI,][]{2012SPIE.8443E..13G,2016SPIE.9905E..1HG} is a non-imaging, soft X-ray telescope aboard the \textit{International Space Station}. The XTI consists of an array of 56 co-aligned concentrator optics (52 currently active) with a field of view of $\sim$30 arcmin$^2$ in the sky. Each unit is associated with a silicon drift detector \citep{2012SPIE.8453E..18P}, operating in the 0.2--12 keV band, yielding a $\sim100$ ns time resolution and spectral resolution of $\sim$85 eV at 1 keV.

For the current study we analysed \nicer data obtained between MJD~59535 and MJD~59605.
Data were reduced using {\tt HEASOFT} version 6.29, \nicer DAS version 2020-04-23\_V007a, and the calibration database (CALDB) version v20210707. 
For the analysis, we selected good time intervals with the {\tt nimaketime} script using standard options. After inspecting the resulting light curves in different bands we identified increased flaring activity due to background contamination. To mitigate the effects of the background we altered some of the standard filtering parameters.
We opted for \textit{ISS} not in the South Atlantic Anomaly region, source elevation $>15^\circ$ above the Earth limb ($>30^\circ$ above the bright Earth), and magnetic cut-off rigidity (\texttt{COR\_SAX}) > 2.0 GeV/c.
Unfortunately due to enhanced background activity during the monitoring period a large fraction of the obtained data -- especially during the lower flux states -- were not useful and were filtered out.
Finally, for timing analysis, we performed barycentric corrections to the event times of arrival using the \texttt{barycorr} tool and the JPL DE405 planetary ephemeris.

Because \nicer is not an imaging instrument, the X-ray background is calculated indirectly. For systems in the direction of the Magellanic Clouds the {\tt 3C50} tool \citep{2022AJ....163..130R} method is optimal \citep[see][]{2021MNRAS.503.6187T}. This approach uses a number of background proxies in the \nicer data to define the basis states of the background database.

\subsubsection{Swift}
X-ray monitoring observations of \sxp have been obtained by the
Neil Gehrels \swift\ Observatory \citep[\swift,][]{2004ApJ...611.1005G} X-ray Telescope \citep[XRT, ][]{2005SSRv..120..165B}.
All archival XRT data were retrieved though the UK \swift science data centre\footnote{\url{http://www.swift.ac.uk/user_objects/}}, and were analysed using standard procedures \citep{2007A&A...469..379E,2009MNRAS.397.1177E}. 

\subsubsection{Chandra}
\sxp was observed by \cxo (obsid:18885) in July 2016 (MJD 57575.34) with an exposure time of 25 ks.
Analysis of these data have revealed the presence of a single peaked pulse profile at $L_{X}{=}$4\ergs{36} in the 0.3--10.0 keV band \citep{2017MNRAS.470.4354V}, 
In the current study we used the same data to create pulse profiles for comparison with the pulse profiles during the 2021 outburst.
Data reduction was performed with the {\tt CIAO} v4.13
software \citep{2006SPIE.6270E..1VF} using standard options through {\tt chandra\_repro} script. Source events were extracted from a 5\arcsec\ region.

\section{Results}
\label{sec:results}

\subsection{Broadband spectral properties}
\label{sec:spec}

\begin{figure}
\begin{center}
\resizebox{\hsize}{!}{
    \includegraphics[angle=0,clip=]{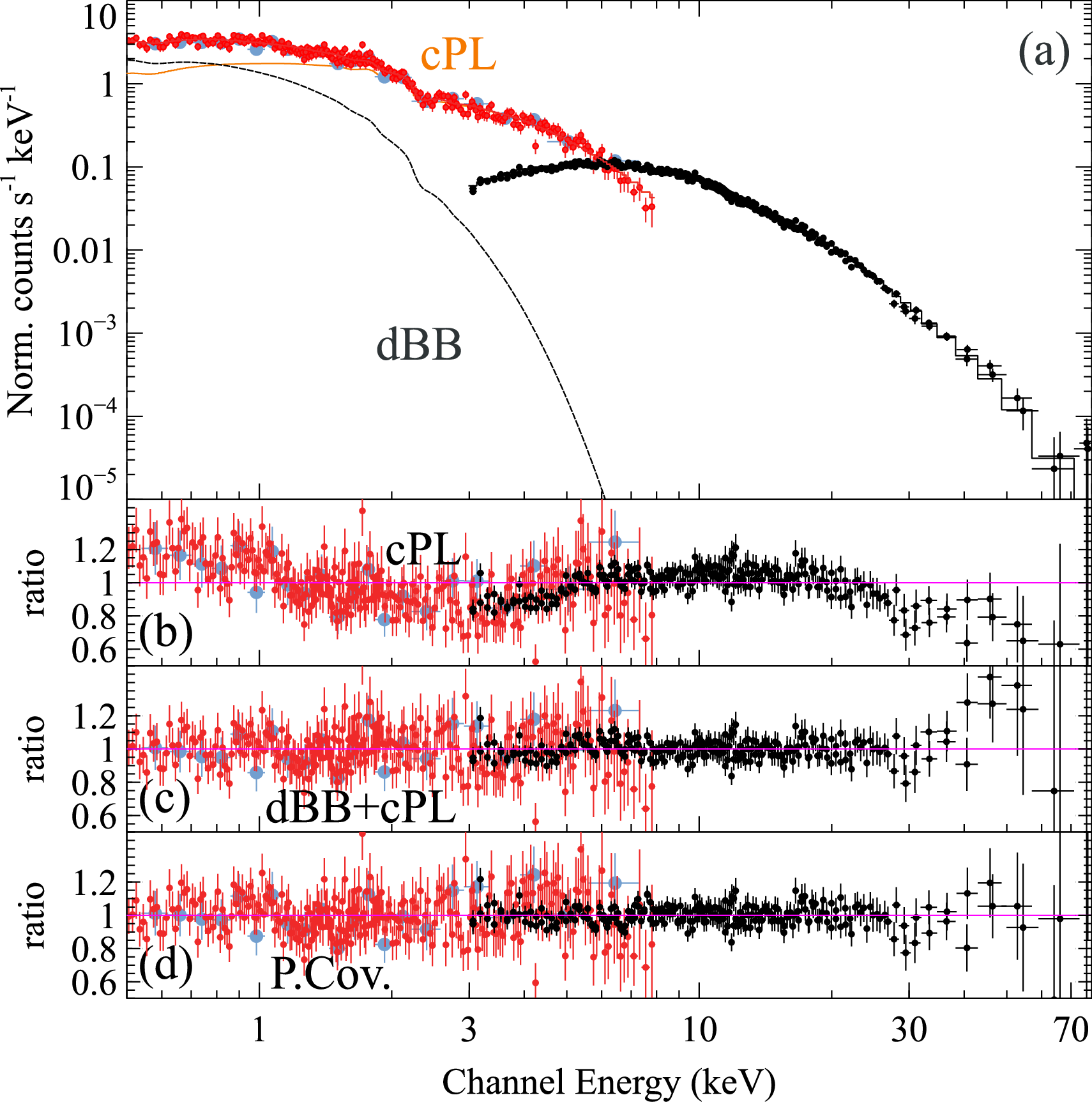}}
\end{center}
  \caption{Broadband spectrum of \sxp. \nustar spectra shown with black colours, while 2 \nicer snapshots are shown with red and cyan colours. 
  Upper panel presents the best fit model composed by an absorbed cut-off power-law and a soft thermal component. 
  Ratio plots show the comparison of the tested models and the data: (b) Absorbed cut-off power-law, (c) Absorbed cut-off power-law plus disk black-body, and (d) cut-off power-law with partial coverage.
  Model parameters are shown in Table \ref{tab:spectral}.}
  \label{fig:broad1}
\end{figure}

Spectral analysis was performed using \xspec v12.10.1f \citep{1996ASPC..101...17A}. 
Photo-electric absorption by the interstellar gas was modelled by {\tt tbabs} component in \xspec, with Solar abundances set according to  \citet{2000ApJ...542..914W} and atomic cross sections from \citet{1996ApJ...465..487V}. 
For fitting the \nicer spectra we used PG-statistics, which implements Cash-statistics \citep{1979ApJ...228..939C}, with a non-Poisson background model. For the modelling of \nustar spectra (3.0--78 keV) we used Cash-statistics. We note that \nus spectra are dominated by background above 40 keV and thus lack the sensitivity to detect faint features at high energies.

It is typical for BeXRBs in the Magellanic Clouds to model the X-ray absorption with a combination of two components to account for Galactic absorption and intrinsic absorption within the SMC and around the binary
\citep[e.g.][]{2013A&A...558A..74V,2016MNRAS.461.1875V,2017MNRAS.470.1971V}. 
The first component was fixed to the Galactic foreground value of 4\hcm{20} \citep{1990ARA&A..28..215D}.
The second component was left free to account for the absorption near the source or within the SMC, thus elemental abundances were fixed at 0.2 Solar \citep{1992ApJ...384..508R}. 
In the spectral modelling we will explore if we see evidence of extra absorption on top of the Galactic one.

In BeXRBs, hard X-ray spectra originate from the accretion column, and can be fitted by a phenomenological power law-like shape with an exponential high-energy cut-off above (or around) 10 keV \citep[e.g.,][]{2013A&A...551A...6M,2014MNRAS.444.3571S,2018MNRAS.474.4432J,2020MNRAS.494.5350V}. In many cases BeXRB spectra show residuals at soft energies. These residuals are often referred to as a ``soft-excess'' with a physical origin that is attributed to one or a combination of mechanisms like emission from the accretion disk, emission from the NS surface, hot plasma around the magnetosphere, or partial absorption from material around the NS \citep{2004ApJ...614..881H}. In this study we investigated the broadband spectrum for the above signatures.

\begin{table*}
\caption{Best-fit parameters spectral empirical models\label{tab:spectral}}
\begin{threeparttable}[b]
\resizebox{\hsize}{!}{
\begin{tabular*}{\textwidth}[t]{llcccccr}
% \begin{tabular*}{\columnwidth}[t]{p{0.18\columnwidth}p{0.2\columnwidth}p{0.2\columnwidth}p{0.4\columnwidth}}
\hline
\hline\noalign{\smallskip} 
% \multicolumn{4}{c}{\xspec: {\tt Tbabs*(diskbb + bbodyrad + cut-offpl)}} \\
Model &  & cPL & cPL+dBB & PC*cPL & PC*(cPL+dBB)  & PC*(cPL+dBB) & \\
\hline
\hline
 & Parameter  &  Value  & & &  &   & Units  \\
\hline  
\hline\noalign{\smallskip}  
 & $N_{\rm{H}}$  Gal$^{(a)}$   & $4$ (fixed) & $4$ (fixed)& $4$ (fixed)& $4$ (fixed)& 4 (fixed)&  $10^{20}$cm$^{-2}$\\\noalign{\smallskip} 
 \\
  & $N_{\rm{H}}$  SMC$^{(a)}$   & $\rightarrow0$  & $\rightarrow0$& $\rightarrow0$ & $\rightarrow0$ & $<3$ &  $10^{20}$cm$^{-2}$\\\noalign{\smallskip} 
 \\
{\tt TBpcf} & $N_{\rm{H}}$ & - & -& $53_{-6}^{+7}$ & $47_{-10}^{+12}$ & $53_{-5}^{+7}$  & $10^{22}$cm$^{-2}$\\\noalign{\smallskip} 
 & {cov. frac.}  & -& -& $45\pm3$&  $32_{-7}^{+8}$& $45\pm3$ &  $\%$ \\\noalign{\smallskip} 
\\
{\tt dBB$^{(b)}$} & k${\rm T_{BB}}$ & - & $0.44_{-0.06}^{+0.014}$ & -& $0.39_{-0.08}^{+0.07}$  & 0.07 (fix) & keV\\\noalign{\smallskip} 
& {$\rm Norm_{BB}$}  &- & $6.3{-0.6}^{+5.3}$  & -&  $6.7_{-2.5}^{+5}$ & $2100_{-1400}^{+23000}$ &   $\rm \sin{\theta}~(R_{BB}/D_{10})^2$ \\\noalign{\smallskip} 
\\
{\tt cPL$^{(c)}$} & ${\rm \Gamma}$  &${0.85}{\pm}{0.02}$ &  $0.49_{-0.03}^{+0.05}$& $1.05\pm0.03$ &   $0.82_{-0.1}^{+0.11}$  & $1.04_{-0.03}^{+0.04}$& - \\\noalign{\smallskip} 
 & ${\rm E_{\rm c}}$ & $20.6\pm0.9$ &  $12.8_{-0.5}^{+0.7}$& $18.6_{-0.9}^{+1.1}$ &   $15.6_{-1.2}^{+1.7}$ & $18.6_{-0.9}^{+1.1}$&  keV\\\noalign{\smallskip} 
 & ${\rm Norm}$  $^{(c)}$  & $101\pm2$ &  $89.8_{-1.6}^{+1.7}$&  $109.4_{-1.8}^{+1.8}$&  $100_{-4}^{+5}$&  $109.4_{-1.8}^{+2.1}$&  \oergcm{-12}\\
\\
\multicolumn{4}{l}{Other information}\\
\multicolumn{2}{l}{Total fit stat. / DOF} &$1399.56/1035$ &$954.11/1033$&  $940.51/1033$&  $923.36/1031$ & $940.47/1032$&  \\
\multicolumn{2}{l}{Goodness $^{(d)}$} &$100\%$ & $4\%$&  $1\%$&  $0\%$ & $4\%$ &  \\
\hline\noalign{\smallskip}
 & $\rm L_{\rm X}$  $^{(e)}$ &$4.67{\pm}{0.08}$   &$4.13{\pm}{0.08}$ & $5.03{\pm}{0.08}$ & ${4.6\pm0.2}$  & ${5.03\pm0.10}$ & \oergs{37}\\\noalign{\smallskip} 
 & $\dot{M}$  $^{(e)}$     &$2.60{\pm}{0.03}$    &$2.30{\pm}{0.04}$ & ${2.80\pm0.04}$  & ${2.55\pm}{0.13}$  &  ${2.8\pm}{0.05}$&  $10^{17}$g s$^{-1}$ \\
  \hline\noalign{\smallskip}  
\end{tabular*}
}
\tnote{(a)} Galactic absorption was fixed to this value (see text for details).
\tnote{(b)} Disk Black-body ({\tt diskbb} in {\tt xspec}) radius may be estimated from the normalisation of the model and distance of 62 kpc (i.e. $D_{10}=6.2$) assuming a disk inclination angle $\theta$.
\tnote{(c)} Cut-off power-law ({\tt cut-offpl} in {\tt xspec}) where normalisation is the unabsorbed X-ray flux in the 0.3--80 keV band.
\tnote{(d)} Based on \xspec simulations, denotes the percentage of simulated spectra that when fitted with the same model yield lower test statistics than the data.
\tnote{(e)} Unabsorbed X-ray luminosity (0.3--80 keV) for a distance of 62 kpc for the cut-off Power-law component. Mass accretion rate onto the NS, assuming $L_{\rm X}=0.2\dot{M}c^2$. 
\end{threeparttable}
\end{table*}

\nustar data were obtained quasi-simultaneously (i.e. less than a day apart) with 2 \nicer visits (obsid: 4202430107 and 4202430108) as seen in Fig. \ref{fig:nicer}.
An absorbed cut-off power-law model sufficiently describes the \nus spectrum. However, fitting the combined \nicer and \nus data we see significant structure in the residuals caused by the soft excess (see Fig. \ref{fig:broad1}).
To eliminate the residual structure we need to add either a partial covering absorber or a soft thermal component to the model (see lower panels in Fig. \ref{fig:broad1}). 
For the thermal component we used a disk black-body ({\tt diskbb} in {\tt XSPEC}).
We also included a model with combined partial coverage on a continuum composed by cut-off power-law and a thermal component. 
In all our tests the column density of the SMC intrinsic absorption was unconstrained and tends to zero.
For the partial covering model, the Tuebingen-Boulder ISM model ({\tt tbpcf} in {\tt xspec}) only provide spectral multiplicative components for standard Solar abundances, as a first order approximation for the SMC abundances $N_{\rm H}$ values obtained by this model should be increased by a factor of $\sim$5 compared to Galactic ones. 
The best fit parameters of all tested models are shown in Table \ref{tab:spectral}. 
Based on the derived parameters we also computed the mass accretion rate corresponding to the bolometric luminosity by assuming that all gravitational energy is converted to radiation\footnote{i.e. $L_{ \rm X}=G\dot{M}M/R \approx 0.2~\dot{M} c^2 (M/1.4 M_\odot)(R/10~{\rm km})^{-1}$} \citep[e.g.][]{2018A&A...610A..46C,2002apa..book.....F}.
We note that for all the tested models the best fit yields an acceptable fit statistics with reduced $\chi^2$ lower than one. 
To estimate uncertainties we implemented a Markov Chain Monte Carlo through \xspec. We used the Goodman-Weare algorithm with 20 walkers and a total length of 50000. For the initial burn-in phase we needed 30000 steps before the chain reached equilibrium. We then generated parameter errors (90\% confidence) based on the chain values.
To further test the goodness of the fit we also simulated spectra based on the MCMC chain parameters. We found that for all models apart from the simplest one (i.e. absorbed cut-off power-law) only a small number of the simulated spectra had a better fit statistics than the real spectra. 
Thus we should be at the limit of our capabilities in testing more complicated spectral models.
We finally note that we found no evidence of an Fe K$_\alpha$ line in the spectrum or any broad absorption feature that is consistent with a CRSF.

\subsection{Long-term light curves and outburst evolution}

\begin{figure}
\begin{center}
\resizebox{\hsize}{!}{
  \includegraphics[angle=0,clip=]{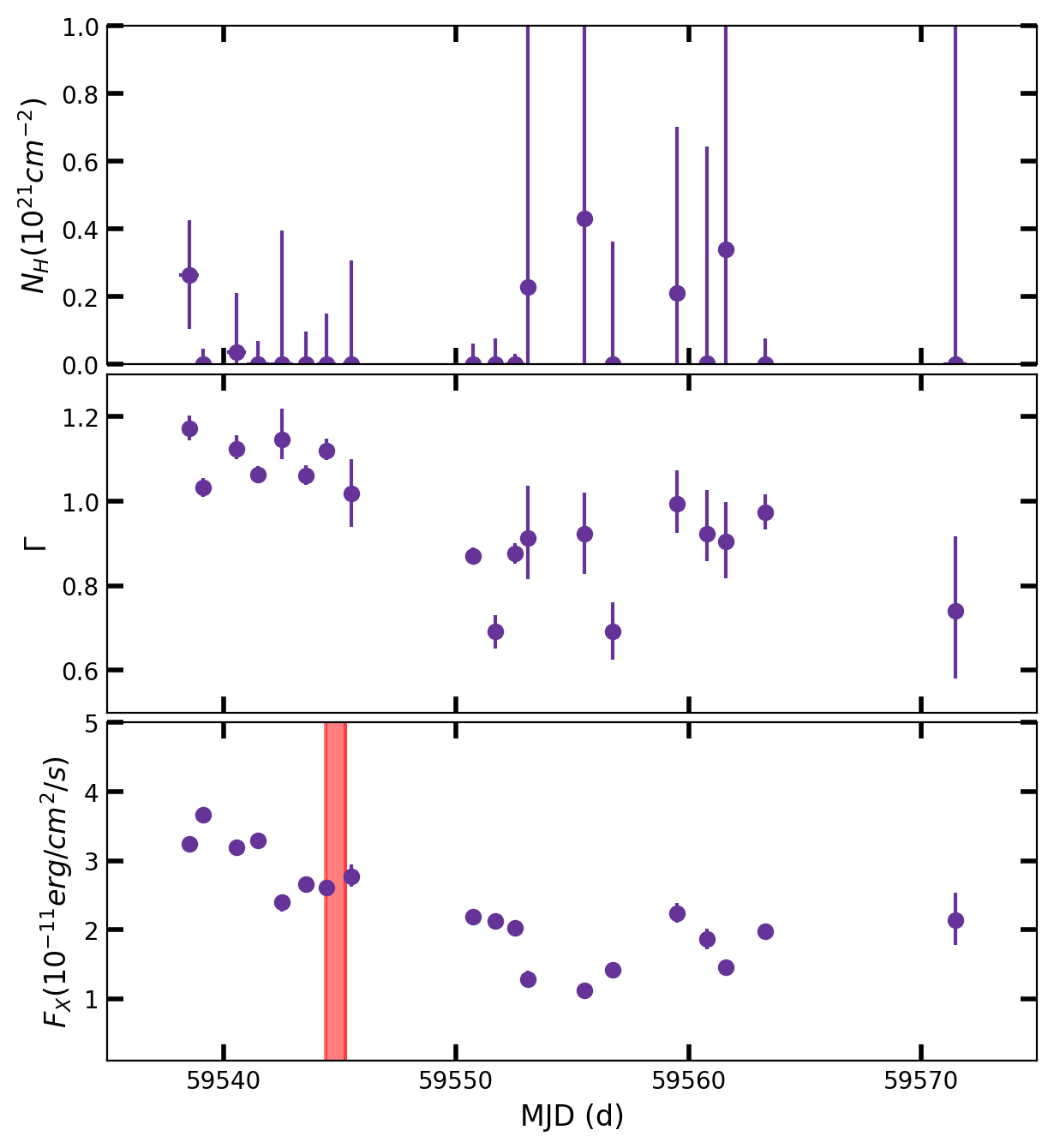}}
\end{center}
\vspace{-0.5cm}
  \caption{Spectral results of \nicer monitoring data obtained during the 2021 outburst of \sxp. 
  The bottom panel shows the unabsorbed flux computed in the 0.3--10.0 keV band. The red-shaded region indicates the epoch of the \nustar ToO.
  No strong evidence for a change in the column density of the absorbing material (top panel) is seen, while we did find evidence of spectral evolution at higher fluxes with the spectral shape (characterised by $\Gamma$, middle panel) becoming softer when brighter above 2\ergcm{-11} (0.3--10.0 keV).}
  \label{fig:nicer}
\end{figure}

\begin{figure}
\begin{center}
\resizebox{\hsize}{!}{
  \includegraphics[angle=0,clip=]{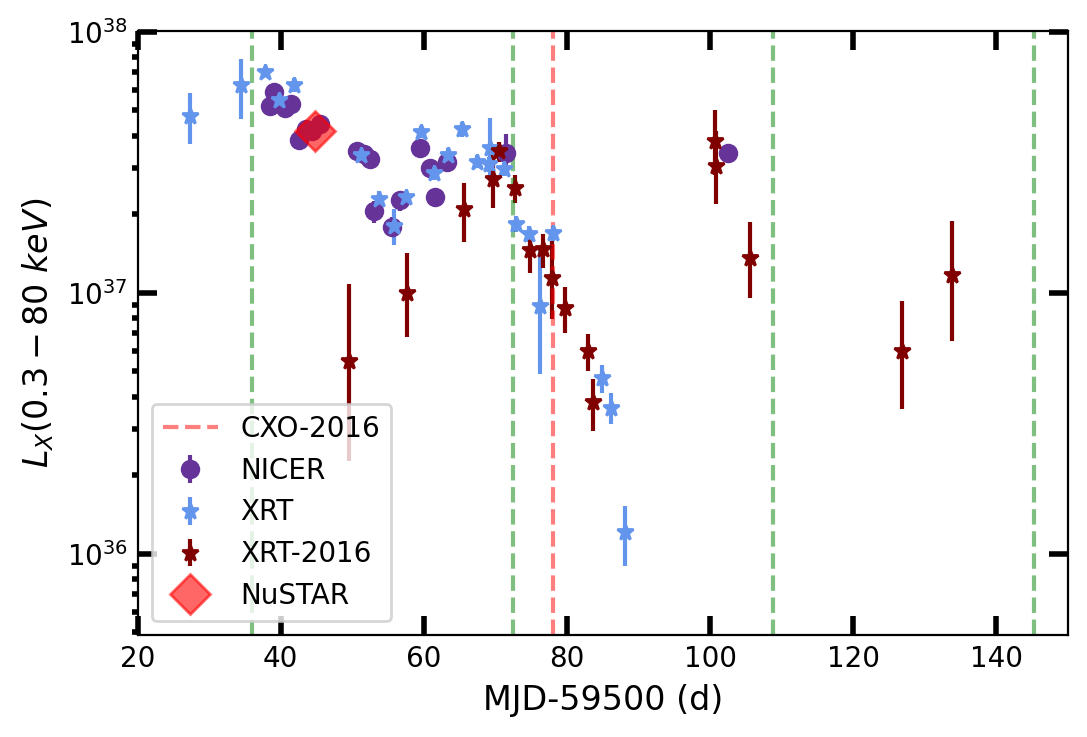}}
\end{center}
\vspace{-0.5cm}
  \caption{X-ray light curves of the 2016 and 2021 outbursts. Luminosities are absorption corrected  (0.3-80.0 keV) and computed for the SMC distance. Archival data from 2016 (first point at MJD 57547) are shifted in time to match the 2021 luminosity peaks. Vertical green lines mark the 36.411 d optical period phased on the first X-ray maximum. For comparison we also show the shifted epoch of the \cxo ToO that falls around the same orbital phase as the \nustar ToO.
 }
  \label{fig:Lx}
\end{figure}

\begin{figure}
\begin{center}
  \resizebox{\hsize}{!}{\includegraphics[angle=0,clip=]{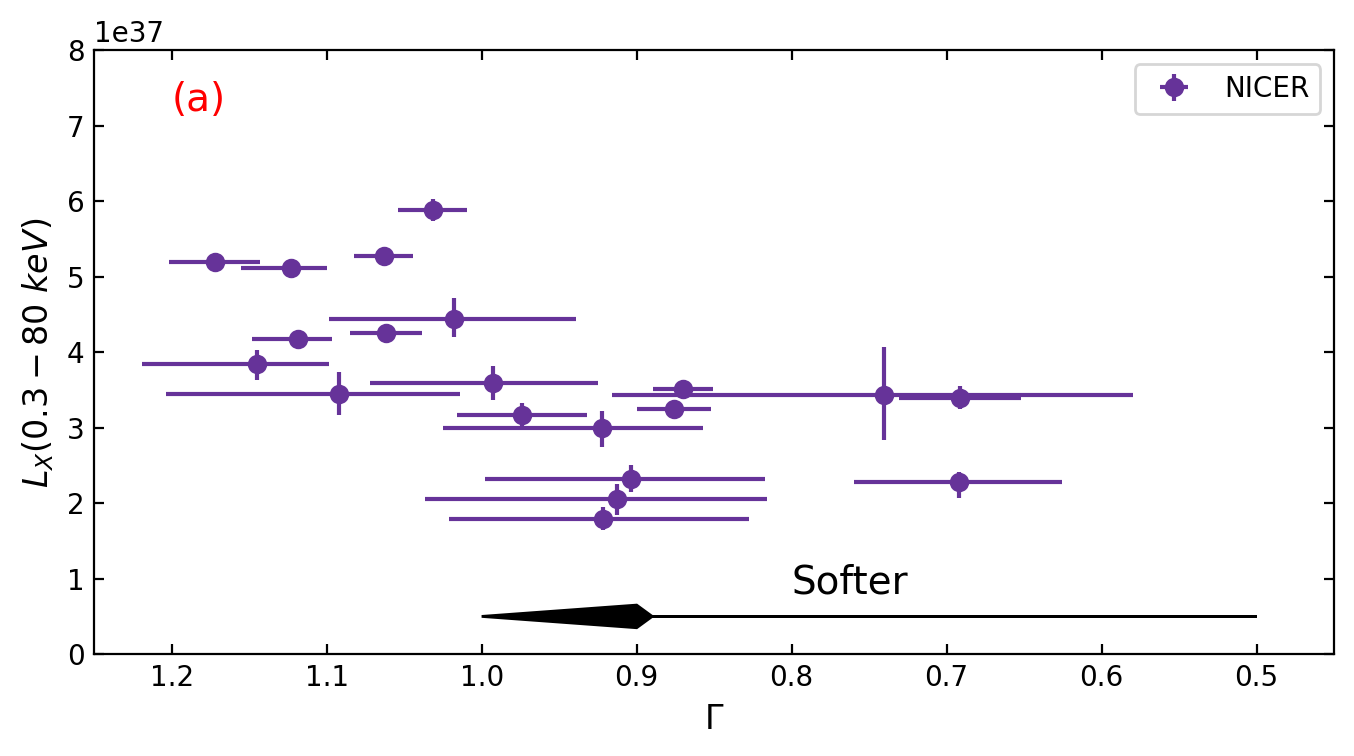}}
  \resizebox{\hsize}{!}{\includegraphics[angle=0,clip=]{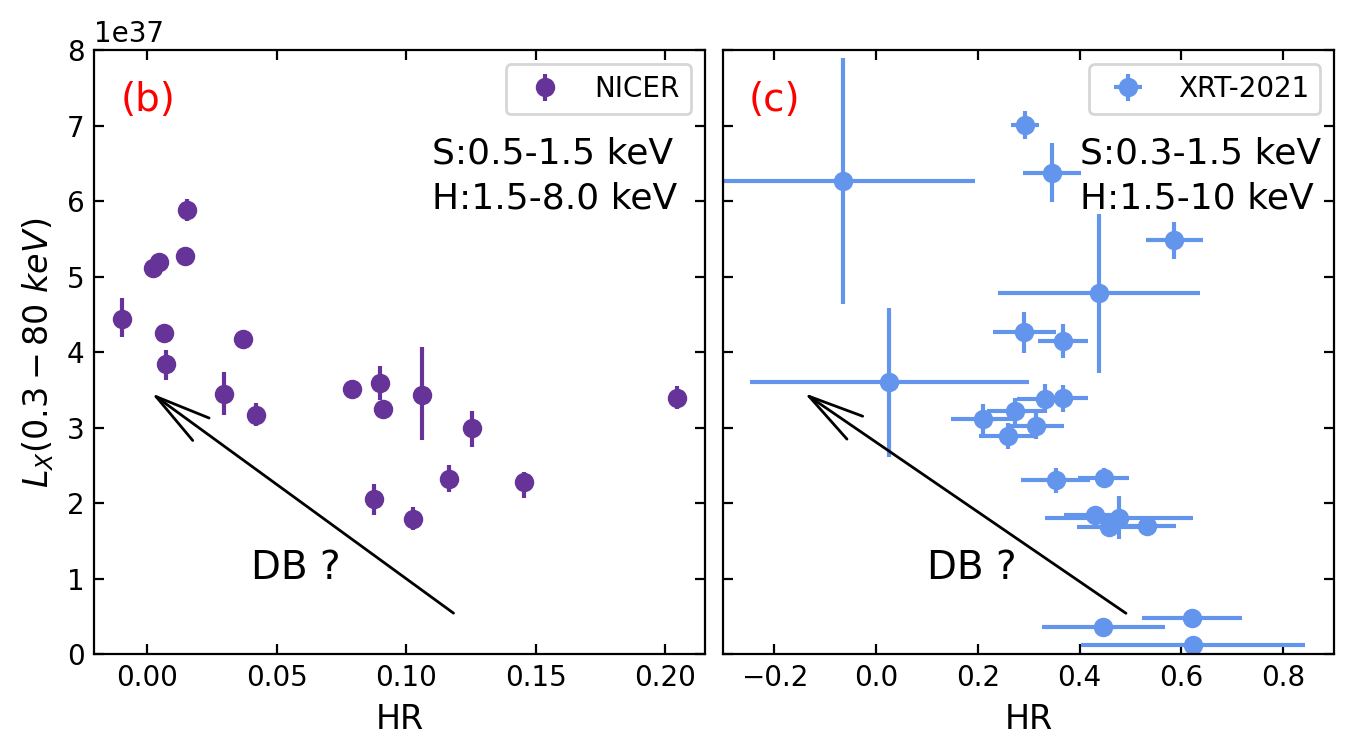}}
\end{center}
\vspace{-0.5cm}
  \caption{Hardness intensity diagram of the 2021 outburst as monitored by \nicer and \swift/XRT.
  The upper panel shows the results of the \nicer spectral analysis, while the lower panels show the results based on hardness ratios (HR).
  There is evidence that the source has entered the diagonal branch (DB) thus the system appears to be close or above the critical regime.}
  \label{fig:HR}
\end{figure}

Monitoring data in soft X-ray band with \nicer enable us to study the evolution of the 2021 outburst as well as to compare it with \cxo data from 2016.
\nicer monitoring data is of sufficient statistical quality to allow us to extract enough counts and perform spectral modelling.
The typical \swift/XRT exposure within one day is of the order of 1000--2000 s, however the effective area of the detector is significantly smaller than that of \nicer. Thus we will use XRT data to only estimate average count rates for each XRT data-set. 

The 20 individual \nicer spectra were fitted with an empirical absorbed power-law model in the 0.5--8 keV band. We also attempted to fit the spectra with a power law with cut-off but the cut-off always converged to very high values (i.e. above 100 keV).
This was not surprising since the cut-off seen in the broad spectra is well above the upper bound of \nicer spectra.
For spectral modelling we used two absorption components as described above.
We found the absorption to be consistent with the fixed foreground value of 4\hcm{20} while the power-law  photon index had a mean value of $\Gamma=0.97$, with evidence of the spectrum becoming softer when brighter at the brightest phase of the outburst.
In Fig.\,\ref{fig:nicer} we plot the evolution of the spectral parameters and the flux for a 40 day interval
(only one observation exists after MJD 59580).

Having analysed the \nicer and \swift/XRT monitoring data as well as the broadband spectra we can compute the evolution of the bolometric luminosity during the 2021 outburst and compare it with archival data. We converted the XRT count rates to 0.3--10.0 keV fluxes using the average spectral parameters inferred from the \nicer spectral fits. A conversion factor of 3.16$\times10^{37}$ erg cm$^{-2}$ s$^{-1}$ / (c/s) was used for all XRT data, while errors are estimated based on count rate uncertainties. The broadband unabsorbed $L_{\rm X}$ was estimated from the broadband spectra. Most of the energy is emitted above 10 keV as the ratio of the broadband (0.3--80 keV) to narrow band (0.3--10.0 keV) luminosity was $\sim$3.5. The 2021 X-ray light curve is shown in Fig. \ref{fig:Lx}. On the same figure we overplot the 2016 XRT monitoring data, time-shifted so the main peak of each outburst match.

Monitoring data can also be used to investigate if the system has entered the super-critical regime, where the accretion column has been formed above the NS surface \citep{2012A&A...544A.123B}. The simplest proxy for this transition is the change of the hardness of the spectrum with intensity \citep{2013A&A...551A...1R}. 
Following the nomenclature of \citet{2013A&A...551A...1R}, for low luminosities the spectrum of many BeXRB pulsars appear to be harder-when-brighter (i.e. so-called horizontal branch) while above a critical limit the systems enter the diagonal branch where they appear to be softer-when-brighter. 
For the intensity we use the bolometric corrected $L_{\rm X}$, while for the colour proxy we use the power-law photon index from \nicer, and the \swift/XRT hardness ratios. We define the hardness ratios (HR) as $\rm{HR}=(\rm{H}-\rm{S})/(\rm{H}+\rm{S})$, where $\rm[H,S]$ is the count rate in a specific hard and soft energy band. 
In Fig. \ref{fig:HR} we plot the intensity-colour diagram of \sxp from 2021 monitoring data.
There is evidence that the system has entered the diagonal branch and appears to be 
close or above the critical limit for accretion column formation.

\begin{figure}
\begin{center}
\resizebox{\hsize}{!}{
  \includegraphics[angle=0,clip=]{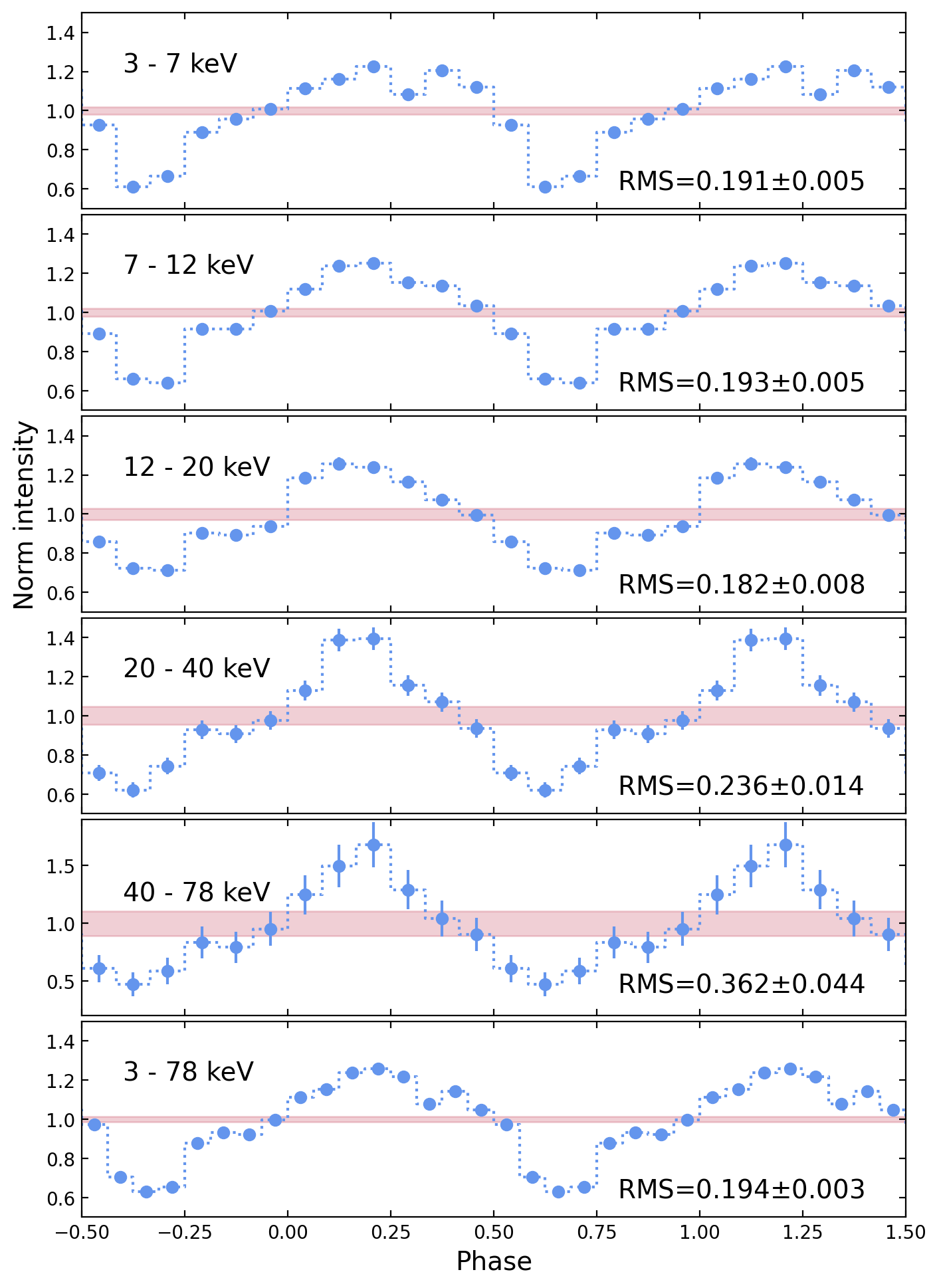}}
\end{center}
\vspace{-0.5cm}
  \caption{Pulse profiles of \sxp from the \nustar observation. Different energy ranges are used to track the changes with energy. Shaded horizontal regions denote the limit of statistical significant variability above the constant level hypothesis.}
  \label{fig:PPnu}
\end{figure}

\begin{table}
\caption{Spin periods \sxp.}
	\label{tab:times}
\begin{threeparttable}[b]
\resizebox{\hsize}{!}{
\begin{tabular}{lcccc} % four columns, alignment for each
\hline\smallskip
Obsid & MJD$\dagger$ (d) & $\Delta{t}$ (d) &  P (s) & \#ToAs$\ddagger$\\
\hline\smallskip
\nicer & & & & \\\smallskip
4202430101 & 59538.50 & 0.8  & 15.6398$\pm$0.0004 & 9 \\\smallskip 
4202430102 & 59539.12 & 0.3  & 15.6394$\pm$0.0019 & 5 \\\smallskip 
4202430103 & 59540.51 & 0.7  & 15.6395$\pm$0.0005 & 6 \\\smallskip 
4202430104 & 59541.44 & 0.9  & 15.6394$\pm$0.0005 & 9 \\\smallskip 
4202430106 & 59543.48 & 0.8  & 15.6393$\pm$0.0009 & 5 \\\smallskip 
4202430107 & 59544.34 & 0.7  & 15.6394$\pm$0.0007 & 5 \\\smallskip 
4202430109 & 59550.70 & 0.5  & 15.6414$\pm$0.0009 & 8 \\\smallskip 
4202430110 & 59551.61 & 0.5  & 15.6424$\pm$0.0013 & 4 \\\smallskip 
4202430111 & 59552.45 & 0.7  & 15.6430$\pm$0.0007 & 5 \\\smallskip 
4202430112 & 59553.04 & 0.1  & 15.6438$\pm$0.0010 & EF \\\smallskip 
4202430114 & 59556.72 & 0.03 & 15.6444$\pm$0.003  & EF \\\smallskip
4202430117 & 59560.77 & 0.1  & 15.650$\pm$0.007  & 3  \\\smallskip
4202430118 & 59561.58 & 0.1  & 15.650$\pm$0.004  & 3  \\\smallskip
4202430119 & 59563.27 & 0.1  & 15.6456$\pm$0.007  & 3  \\\smallskip
4202430121 & 59571.45 & 0.9  & 15.6378$\pm$0.0006 & 9  \\\smallskip
\nustar & & & &\\\smallskip
90701339002 & 59544.8 &0.8 & 15.6395$\pm$0.0004 & 16  \\
	\hline
	\end{tabular}
}
\tnote{$\dagger$} Middle epoch of observation. 
\tnote{$\ddagger$} Number of ToAs used for spin period refinement. For those observations not possible we used periods based on epoch folding. 
\end{threeparttable}
\end{table}

\subsection{Temporal properties - pulse profiles}
\label{sec:temp}

To search for a periodic signal we used the epoch folding Z-search method \citep{1983A&A...128..245B} implemented through {\tt HENdrics} command-line scripts and {\tt Stingray} \citep[][]{2019ApJ...881...39H}.
For \nustar data we search for a periodic signal in the 3--40.0 keV range (35-960 PI channel). 
Our final estimate of the spin period and its uncertainties was based on the time of arrival (ToA) method \citep[e.g.][]{2020A&A...637A..33T}.
We first used {\tt HENdrics} to derive a most probable period, then we estimated ToAs of individual pulses for 16 intervals, and we finally used PINT\footnote{\url{https://github.com/nanograv/pint/}} \citep{2021ApJ...911...45L}. From the above
we derived a period of 15.6395$\pm$0.0004 s for the 2021 \nustar data.
The reported period for the 2016 data was  15.6398$\pm$0.0009 s, and is consistent within uncertainties with the new derived period.
For consistency, we implemented the ToA procedure to estimate a period for the 2016 \cxo data. This yielded a period of 15.6396$\pm$0.0014 s. All tests indicated that the period of the NS has remained unchanged within uncertainties for more than 5 years.

The strength of the periodic modulation is typically quantified through the root-mean-squared (rms) pulsed fraction. This is given by:
\begin{equation}
PF_{\rm RMS}=\frac{\left(\sum_{j=1}^{N} (R_{j} - \bar{R})^2/N\right)^{1/2}}{\bar{R}},
\label{eq1}
\end{equation}
where N is the number of phase bins, $R_{j}$ is the background subtracted count rate in the $j^{\rm th}$ phase bin, and $\bar{R}$ is the average count rate in all bins \citep[e.g.][]{2018ApJ...863....9W}. We used the above definition to estimate the $PF_{\rm RMS}$ in different \nustar energy bands. 
We found no significant change in the PF within different energy bands up to 20 keV. However, above 20 keV PF increases and pulsations become almost twice as strong at the highest energies. We also note that at above 40 keV background photons contribution is  $\sim$50\% of the net counts.
The increasing PF with energy is typical for accreting pulsars and is attributed to hard photons that are emitted from the sides of the accretion column being more beamed compared to soft photons \citep[e.g.][]{2009AstL...35..433L}.

In Fig. \ref{fig:PPnu} we present the folded pulse profiles for different energy bands covering the full \nustar energy range. We opted to also show the soft energy band (1.6--5.0 keV) in order to compare with \nicer and \cxo pulse profiles.
In Fig. \ref{fig:PPs} we present the folded pulse profiles from all \nicer observations where a period could be estimated. 

\subsection{Spin evolution}
\label{sec:spin}

To investigate the spin evolution during the outburst we implemented the same method on the \nicer data between 0.5--8.0 keV. 
We first computed the most probable period by epoch folding and then we refined the period and its uncertainty based on the ToA of individual pulses. Given the shape of the pulse profile, the template used for ToAs can drastically vary from observation to observation. 
Thus, we used one universal template for all observations, aiming in characterisation of the off-phase of the pulse. The method was successful when two conditions occurred. Firstly, we need more than three \nicer snapshots to be performed within one day, and secondly the total number of counts must be high enough to obtain meaningful pulse profiles.
 For all other snapshot phases connecting the ToAs was challenging to impossible due to multiple peaks in the periodogram with similar intensity, a problem often encountered in slow pulsars observed with gaps \citep[e.g.][]{2017ApJ...839..125Z,2018A&A...620L..12V}.

\begin{figure}
\begin{center}
  \resizebox{\hsize}{!}{\includegraphics[angle=0,clip=]{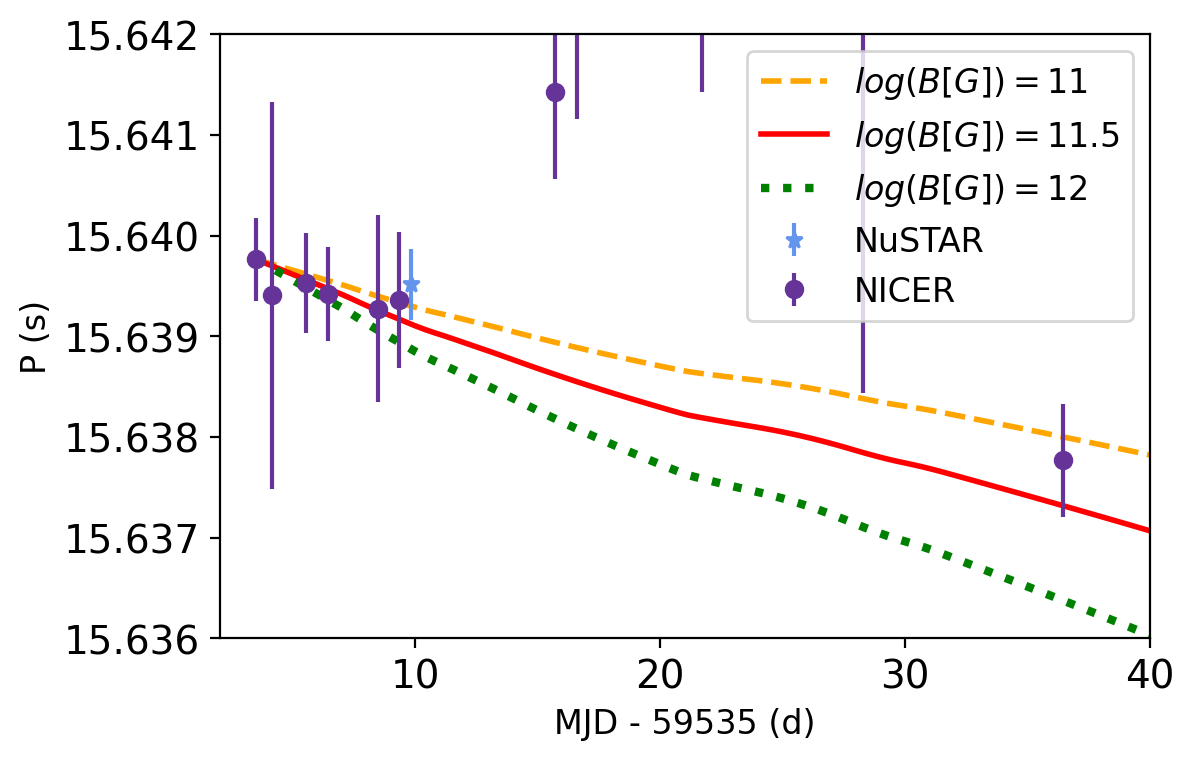}}
\end{center}
\vspace{-0.5cm}
  \caption{Period evolution of \sxp within the 2021 outburst. With dashed lines we plot the predicted spin-up rate based on the observed luminosity and three values of the magnetic field. 
  }
  \label{fig:Pevol}
\end{figure}

The period evolution is shown in Fig. \ref{fig:Pevol}. The overall trend seems linear although a large gap occurs in the data. Although due to the sampling, variability due to orbital Doppler shifts is visible, the span of the \nicer points with good timing solutions is $\sim$33 d and is comparable to the $\sim$36.4 d optical period. Thus the secular evolution between the first and last point only should not be affected much by orbital effects.
With that assumption we found an average spin-up of $\dot{P}=(7.0\pm2.5)\times10^{-10}$ s s$^{-1}$ (or $\dot{\nu}=(2.8\pm1.0)\times10^{-12}$ Hz s$^{-1}$), this value should be the approximate intrinsic spin-up due to accretion. 
Alternatively one can calculate the intrinsic spin up of the NS \citep[see][for method]{2019MNRAS.488.5225V,2020MNRAS.494.5350V} due to mass accretion rate as derived by the observed bolometric $L_X$ (see Fig. \ref{fig:Lx}). 

Here are the basic steps for our calculation. 
We assume mass transfer from a Keplerian disk thus the induced torque due to accretion only is 
$N_{\rm acc}\approx\dot{M}\sqrt{GM_{\rm NS}R_{\rm M}}$. 
The total torque can be expressed in the form of $N_{\rm tot}=n(\omega_\mathrm{fast})N_{\rm acc}$ where $n(\omega_\mathrm{fast})$ is a dimensionless function that accounts for the coupling of the magnetic field lines to the accretion disk \citep[for details see][]{1995ApJ...449L.153W,2016ApJ...822...33P}.
The spin-up rate of the NS is then given by: 
\begin{equation}
\dot{v}=\frac{n(\omega_\mathrm{fast})}{{\rm 2 \pi} I_{\rm NS}} \dot{M} \sqrt{G M_{\rm NS} R_{\rm M}},
\label{eq:torque}
\end{equation}
where $I_{\rm NS} \simeq (1-1.7)\times10^{45}$~g cm$^{2}$ is the moment of inertia of the NS \citep[e.g.,][]{2015PhRvC..91a5804S}.

To model the intrinsic spin evolution due to accretion we just need to numerically solve eq.~(\ref{eq:torque}) in time assuming a constant magnetic field strength. 
We used 1000 time steps between that span over the \nicer monitoring.
For each time step the mass accretion is estimated by interpolating the observed flux in the 0.3-10 keV (see Fig. \ref{fig:nicer}). Then we converted the flux to bolometric luminosity assuming the spectral parameters of the broadband spectra.
Bolometric luminosity was then used as a proxy for mass accretion (i.e. $L_{ \rm X}=G\dot{M}M/R$).
For $n(\omega_\mathrm{fast})$ we follow the \citet{1995ApJ...449L.153W} model (see their eq. 19). For all calculations we adopted standard NS parameters (i.e. 12 km radius, 1.4$M_\odot$ mass and $I_{\rm NS}=1.3\times10^{45}$~g cm$^{2}$).
We repeated this process for various magnetic field strengths, while the results are shown in  Fig. \ref{fig:Pevol}.
It is evident that the observed secular spin up is consistent with a magnetic field strength close to 3$\times10^{11}$~G, while very low (<$10^{11}$~G) or high (>$10^{12}$~G) $B$ values seem to be inconsistent with observations.
Given that the first and last pointing are separated by about one orbital period we can neglect any orbital effects in this first order approximation. However, we will investigate any effects in the next section.

\begin{figure}
\begin{center}
  \resizebox{\hsize}{!}{\includegraphics[angle=0,clip=]{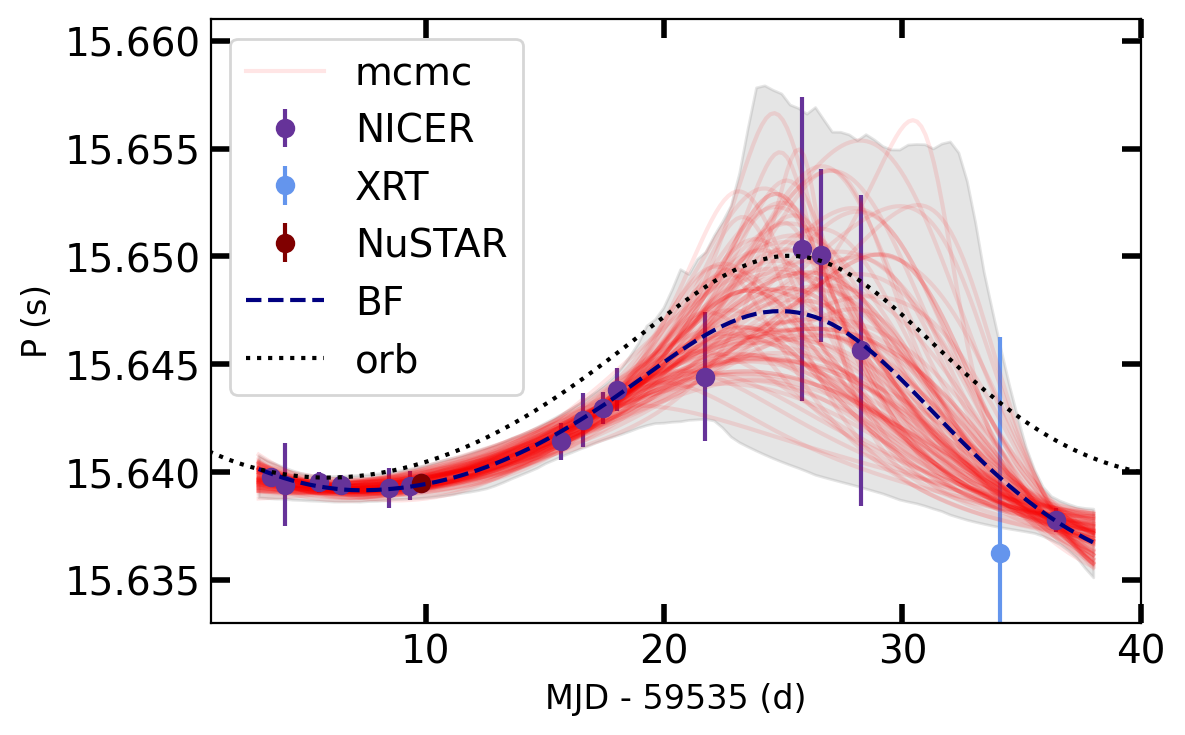}}
\end{center}
\vspace{-0.5cm}
  \caption{
    Modelling of the spin period evolution of \sxp based on orbital modulation and intrinsic spin-up. 
  With the dashed line we show the most probable solution while the black dotted line shows the Doppler shifts due to orbital modulation alone, without the spin-up due to accretion.
  With red lines we show a family of orbits drawn from the sample of the posterior distribution of parameters from the Bayesian modelling. Shaded region is the 99\% quantile of the models. 
  Although \swift/XRT data present hits of periodic modulation the uncertainties are quite high and cannot constrain our model. For visualisation purposes we plot one of those XRT measurements.}
  \label{fig:Porb}
\end{figure}

\begin{table}
	\centering
	\small
    \setlength\tabcolsep{2pt}
	\caption{Orbital parameters of \sxp binary.}
	\label{tab:Orbitals}
	\begin{tabular}{lccr} 
		\hline\smallskip
		Parameter & Prior & Result & Units\\
		\hline\smallskip
		$P_{\rm orb}$ & 36.411 (fixed) & &d\\\smallskip
		$e$ & 0-0.99 & 0.29$_{-0.16}^{+0.15}$  &-\\\smallskip
		$\omega$ & 0-360 & 20$_{-27}^{+26}$ &$^{o}$\\\smallskip
		$a \sin i$ &50-400& 162$_{-40}^{+50}$ & 1-sec\\\smallskip
		$T_{\rm \pi/2}$ & 59520-59560 & 59537.9$_{-1.2}^{+1.2}$ & MJD (d)\\
		\hline
	\end{tabular}
    \label{tab:orbit}
\end{table}

\subsection{Orbital evolution}

The observed spin evolution shows significant variation from the expected evolution due to intrinsic spin up due to accretion. The remaining residuals may be due to orbital modulation. 
Keplerian orbits are described with five orbital elements: orbital period ($P_{\rm orb}$), the orbital eccentricity ($e$),
the argument of periastron of the star’s orbit ($\omega$),  the velocity semi-projected axis ($a\sin{i}$ in light-sec), and finally for the orbital phase we use the time of a mean longitude of 90 degrees (i.e. $T_{\rm \pi/2}$).
Modelling the orbital modulation with the intrinsic spin-up due to accretion that we described in the previous section (see \S \ref{sec:spin}) is often done for \fermi/GBM pulsars \citep{2017PASJ...69..100S}. However, most of the GBM monitored pulsars are Galactic sources which are monitored for extensive periods.
To model our data set it requires to properly map the parameter space in order to identify degeneracies between model parameters.
Thus, to fit the model to the data we implement a nested sampling algorithm for Bayesian parameter estimation and estimate posterior distributions for the parameters of standard accretion torque models and binary orbital parameters. In terms of statistical treatment similar methods have been used to model radial velocity curves from binary systems \citep{2018PASP..130d4504F}.

To derive the posterior probability distributions and the Bayesian evidence we used the nested sampling Monte Carlo algorithm
MLFriends \citep{2004AIPC..735..395S,2019PASP..131j8005B} that employs the
{\sc ultranest}\footnote{\url{https://johannesbuchner.github.io/UltraNest/}} package \citep{2021JOSS....6.3001B}. 
An outline of the above method with applications to accreting pulsars will be presented by \citet{kar22}.

Due to the gaps in the \nicer monitoring and the high background, good timing data exist only for the first 20 days of the monitoring. In an effort to improve our data-set we also searched for pulsations in \swift/XRT data. However, due to low statistics and small number of XRT snapshots within a day typical period uncertainties were of the order of 0.01-0.05 s (see Fig. \ref{fig:Porb}), thus offer little information for our study. 
To constrain our model parameters we fixed the orbital period to 36.411 days and limited the magnetic field strength (i.e. $\log{(B [G])}\in[10,13]$).
With the above assumptions, and for typical parameters of the NS (12 km and 1.4$M_\odot$) we estimated the posterior distribution for our other model parameters.
These parameters are listed in Table \ref{tab:orbit}, while in Fig. \ref{fig:corner} we show the corner plot of posterior distributions for our solution. In Fig. \ref{fig:Porb} we show a sample of 100 random orbital models from the posterior distribution together with the most probable model.

\section{Discussion}\label{sec:discussion}

The latest outburst of \sxp started in late November 2021 while the system remained in a bright state until early 2022. The bolometric luminosity during the event reached 7\ergs{37}, which presents the brightest stage ever observed for the system. 
This luminosity translates to  $\sim30\%$ of the Eddington limit for a typical NS (i.e. 12 km and 1.4$M_\odot$).

The combined \swift/XRT and \nicer light curve of the 2021 outburst reveals a complex structure with two peaks separated by the orbital period. The increased flux at later epochs indicate a third peak, but not enough monitoring data were collected to further explore this behaviour. The general structure of the event matches very well the behaviour seen in 2016 (see Fig. \ref{fig:Lx}), with minor differences in the relative intensity of the three peaks. The mismatch of the X-ray and optical period could be attributed to a precessing decretion disk around the Be star \citep[for an  observational example and a theoretical application see][]{2021MNRAS.503.6187T,2021ApJ...922L..37M}. Such precession can cause an evolving period between outbursts if the disk is moving retrograde to the NS orbit. Nevertheless the self-similarity of the 2016 and 2021 outbursts with three peaks is quite intriguing revealing that the Be disk and the NS geometrical configuration behaves in a repetitive manner.

The NS spin evolution shows evidence of orbital and intrinsic spin-up due to accretion.
Because the first and last day of the \nicer observations (MJD 59535-75 interval) consisted of a large number of snapshots, we were able to constrain the intrinsic spin-up and found a value of $\dot{\nu}=(2.8\pm1.0)\times10^{-12}$ Hz s$^{-1}$.
To improve this estimate, we modelled simultaneously the orbital modulation and intrinsic spin-up with a Bayesian approach.
We found that the spin-up is consistent with a 
NS pulsar with a magnetic field strength of 5$\times$10$^{11}$~G with a factor of $\sim$3 uncertainty.
Moreover, the orbit has moderate eccentricity with a value of $\sim0.29$ while values up to 0.6 cannot be statistically excluded. Interestingly, looking at the corner plot in Fig. \ref{fig:corner} it seems that lower magnetic fields are favoured by more eccentric orbits and the uncertainties in some parameters are still high. 
From the orbital parameters (i.e. $\omega$ and $T_{\rm \pi/2}$) we can also estimate the epoch of periastron $T_{\rm per}$, which is found to be MJD 59531$\pm$4.
The value of $T_{\rm per}$ seems to match or lead by a few days the X-ray maxima observed in the X-ray light curve (see Fig \ref{fig:Lx}). 
Figure \ref{fig:Porb} interestingly shows that quite a few orbits with higher eccentricity are still statistically acceptable and cannot be excluded without better data.
Future independent measurements that could constrain the orbital parameters would help us revisit the system and tighten the constraints on the magnetic field strength.

The value of the magnetic field that we estimate can be compared with the estimates of \citet{2022MNRAS.513.5567C}, who found a value of $3.7\times10^{12}$~G, assuming the source is in spin equilibrium. For a NS pulsar to rotate near spin-equilibrium one assumes constant mass accretion and requires the disk inner radius to rotate with about the same angular velocity as the NS. However, there are a few caveats in such assumptions. Given that BeXRBs are extremely variable systems the assumption of steady accretion does not hold, and for systems rotating near equilibrium torque transfer is a non-linear problem. Moreover, rotating near equilibrium does not mean that the spin-up rate is zero. In fact under the assumption of steady accretion, even some of the systems with the highest spin-up rate can be argued to be near equilibrium \citep{2022MNRAS.513.6219P}. In addition, finding the secular spin evolution to be very small does not necessarily mean that the NS is in equilibrium, the spin evolution is a dynamical problem, where the accretion duty cycle plays an equally important role in determining any secular spin change between two epochs. Thus estimates of the magnetic field assuming spin-equilibrium are subjected to large systematic uncertainties and should be considered much less accurate than our estimates from torque modelling.   

The broadband spectra from the combined \nustar and \nicer observations lack any significant line features that could be associated with emission from hot plasma (like an Fe K$\alpha$ line) or a CRSF. 
The high energy part of the spectrum is well explained by a power law with a cut-off in agreement with other BeXRBs.
However, the soft part of the spectrum requires additional components in order to be explained. 
We tested whether the addition of a thermal component or a partial absorber could improve the quality of the fit. 
Either or both of these components would result in an acceptable fit. 

Of particular interest is the thermal component as it can be interpreted as the inner region of an accretion disk.
An analytical form for the magnetospheric radius can be expressed in the following form \citep{2002apa..book.....F,2018A&A...610A..46C}:
\begin{equation}
R_{\rm M} = 0.5\left(\frac{\mu^{4}}{2GM \dot{M}^{2}}\right)^{1/7}, 
\label{eq2}
\end{equation}
where $M$ is the NS mass, $\mu=B R^{3}/2$ is the magnetic dipole moment, with $R$ the NS radius and $B$ the NS magnetic field strength at the magnetic poles. For typical parameters of the NS (12 km and 1.4$M_\odot$) and for the observed mass accretion rate and assuming a polar magnetic field strength of 5$\times$10$^{11}$~G, the disk radius should be of the order of 600 km. 
However, our best fit spectral model yields a much smaller inner disk radius (i.e. $\sim15-20$ km, see Table \ref{tab:spectral}) which in fact is comparable to the NS radius. Such radius would translate to a B value of 10$^{9}$~G, which is unrealistically small for a BeXRB pulsar. 
This value is also at odds with the results of the torque modelling, as in Fig. \ref{fig:Porb} we see that a magnetic field lower than $10^{11}$~G would not be able to explain the spin evolution over the 30 day period, unless a quite eccentric orbit is assumed.
Since the disk could be seen edge on, another way to estimate its size is from its temperature. By adopting the mass accretion rate from the pulsed continuum, and for a temperature of 0.4 keV (from the fit) the inner radius of a standard disk would be around 60--70 km, depending on the spectral hardening parameter \citep[see][]{1998PASJ...50..667K,2005ApJ...618..832Z}. 
This is still significantly smaller than the size estimated from torque modelling, which predicts a disk size of about 600 km for a magnetic field of $5\times10^{11}$G and $\dot{M}\sim2.5\times10^{17}$~g~s$^{-1}$. For such parameters the disk temperature should be  $\sim$70 eV.
As seen in the last column of Table \ref{tab:spectral} a model with a disk with fixed temperature at 70 eV, can sufficiently explain the observed spectra. In fact we see that the disk size for such temperature is of the order of 200--900 km.

With respect to the partial coverage model, the rationale behind this model is that the line of sight between the observer and the source is not a line in a mathematical sense. Given the extent of the emitting region, the geometrical problem is better described as a superposition of multiple line of sights. The absorber present in the vicinity of the source can be imagined as a collection of dense clouds, that lie inside the magnetosphere or engulf the binary itself. Partial covering manifests as broad humps or bumps below 10 keV depending on the column density of the partial absorber. 
For example for the Galactic pulsars GX 304-1 and Her X-1 the partial absorber has typical values of $N_{\rm H}\sim$10-70\hcm{22} and covering fraction of 30-40\% depending on the selection of the continuum model \citep[e.g.][]{2000PASJ...52..223E,2014PASJ...66...44A,2016MNRAS.457.2749J}. These values are quite similar to the values we find in \sxp and other BeXRB systems in the MCs \citep[e.g.][]{2018MNRAS.475..220V}.  

The above discussion demonstrates the complexity of the soft excess which in our case is not possible to model without over-fitting the data. Thus we caution the reader on the spectral parameters for the soft excess. For these empirical models a physical interpretation is difficult since one can get acceptable goodness of fit and almost flat spectral residuals for a wide range of parameters.

We can use the results of the broadband spectroscopy to convert observed fluxes in the 0.3-10.0 keV band to bolometric luminosities. 
The bolometric correction is $\sim$3.5--4, where its uncertainty is due to the model selection, since as partial covering models generally give higher unabsorbed flux. 
This is quite high for an accreting pulsar, where it is typically expected that 40-50\% of the energy is released in the 0.5--10.0 keV band \citep{2022MNRAS.513.1400A}. This demonstrates the quite large variety of spectral shapes in XRPs that can be lost if average bolometric corrections are adopted from a small sample size \citep[i.e. 9 pulsars in][]{2022MNRAS.513.1400A}.
The bolometric correction is important if we would like to track transitions in the spectral properties associated with changes in the accretion column \citep{2012A&A...544A.123B}.
Following \citet{2013A&A...551A...1R} we can argue that the spectral behaviour during the 2021 monitoring is consistent with the super-critical regime, thus we are above the critical luminosity where the accretion column has started to form \citep[see also][]{2015MNRAS.452.1601P}.
Following \citet{2012A&A...544A.123B} this luminosity is given by:
\begin{equation}
L_{\rm crit}=\left(\frac{B}{0.688\times10^{12}~G}\right)^{16/15}{\times}10^{37}~{\rm erg\,s^{-1}},
\label{eq3}
\end{equation}
that holds for typical parameters \citep[see eq. 32 of][for more details]{2012A&A...544A.123B}.
From the colour intensity diagram (see Fig. \ref{fig:HR}) we can set an upper limit to the critical transition and thus the magnetic field of the NS. For $L_{\rm crit}$ of the order of $\sim$\oergs{37} we found an upper limit for the magnetic field of the order of 0.7$\times$10$^{12}$ G. 
This quantitative estimate seems to be in agreement with the lack of any evident CRSF in the broadband spectrum.
Such a feature would appear at energies ($E_{\rm CRSF}$) related to the magnetic field through the following relation:
\begin{equation}
B_{\rm CRSF}=(1+z)\left(\frac{E_{\rm CRSF}}{11.57\,\rm keV}\right){\times}10^{12}~{\rm G},
\label{eq4}
\end{equation}
where $z\sim0.3$ is the gravitational red shift from the NS which is related to the NS compactness \citep[e.g.][]{2019ApJ...887L..21R}. The upper limit for the critical luminosity would thus translate to an upper limit for the $E_{\rm CRSF}$ of $\sim$6 keV. Given the complexity of the spectrum at lower energies and the presence of the soft excess it is thus not surprising that we did not detect any CRSF in the spectrum. This is also consistent with the fact that no accreting pulsar has a detected electron CRSF below 8 keV, and claims of such detections in some systems still wait to be confirmed \citep[][]{2019A&A...622A..61S}.

The pulse shape of the pulse profile is single peaked (Fig. \ref{fig:PPnu} \& \ref{fig:PPs}). At higher energies ($>40$ keV) the pulse profile is sharper with a triangular shape, while it appears to be broader in the soft band with weak evidence of a secondary peak. 
We note the similarity of the pulse profile in the soft \nustar band (see Fig. \ref{fig:PPnu}) to those obtained from the \nicer monitoring (see Fig. \ref{fig:PPs}). The presence of a single peaked pulse profile at what otherwise 
appears to be close or above the critical limit for the super-critical regime is quite unusual as complex profiles are the norm for such bright BeXRBs \citep[e.g.][]{2017MNRAS.472.3455E,2018A&A...614A..23K}. Further investigation of the pulse profiles with physical models \citep[e.g.][]{2017PASP..129l4201C,2018MNRAS.474.5425M} could provide further information of this somewhat puzzling feature. 

\section{Conclusions}

We have analysed broadband spectra from the 2021 outburst of \sxp. We did not identity any CRSF that could provide a direct measurement of the NS magnetic field. 
Nevertheless, the lack of such a feature does not exclude its presence as a CRSF could also be quite weak or pulse-phase dependent \citep[e.g.][]{2013Natur.500..312T}.
In the broadband spectra, we found evidence of a soft spectral component that could be associated with an accretion disk, however its parameters are not well constrained. An alternative explanation is that the soft-excess is a result of partial absorption, and we do not favour one model over the other. 
The evolution of the spectral properties during the 2021 outburst is consistent with the presence of an accretion column above the NS, and the system accreting close or above the critical limit. Finally, we did not measure any secular evolution of the spin period of the pulsar between 2016 \cxo and 2021 \nustar observations, which is consistent with the findings of \citet{2022MNRAS.513.5567C}.
Nevertheless, there is evident modulation in the \nicer monitoring data that is consistent with orbital motion of the binary. Modelling of the orbital motion and intrinsic spin-up during the outburst enabled us to constrain the magnetic field strength and the orbital parameters of the system.
All the derived quantitative and qualitative results consistently provide indirect constraints on the NS magnetic field strength. Given the lack of a cyclotron line the most reliable measurement comes from the intrinsic spin-up due from which we find a value of 5$\times$10$^{11}$~G with an uncertainty of the order of a factor of $\sim$3.

\begin{acknowledgements}
The authors would like to thank the anonymous referee for the constructive report that helped to greatly improve the manuscript.
GV would like to thank P.S. Ray for his comments, suggestions and advice on using PINT. 
This work was supported by NASA through the \nicer mission and the 
Astrophysics Explorers Program.
Facilities: \nicer, \nustar.
We acknowledge the use of public data from the \swift data archive.
\end{acknowledgements}

\bibliographystyle{aa} % style aa.bst
\bibliography{general} % your references Yourfile.bib

% Appendix

\begin{appendix}
% \onecolumn

\section{Extra figures and Tables}

\begin{table*}
\centering
\caption{Power-law fit to \nicer data\label{tab:spectral_nicer}}
\begin{threeparttable}[b]
% \resizebox{\hsize}{!}{
\begin{tabular}{ccccccc}
% \begin{tabular*}{\textwidth}[t]{ccccccc}
% \begin{tabular*}{\columnwidth}[t]{p{0.18\columnwidth}p{0.2\columnwidth}p{0.2\columnwidth}p{0.4\columnwidth}}
\hline
\hline\noalign{\smallskip} 
OBSID      & MJD	 &  N$_{\rm H}$ $^{a}$	& $\Gamma$ &  $F_X$ (0.3-10.0 keV)&  $\chi^{2}_{\rm red}$ &	DOF     \\ 
--      & d	 &  \hcm{22}	& -- & \oergcm{-11}&  -- &	--     \\ 
\hline
4202430101 & 59538.5 & 0.0263$^{+0.016}_{-0.016} $ & 1.17$^{+0.03}_{-0.03} $ & 3.24$^{+0.07}_{-0.07}  $ & 1.15 & 424 \\\noalign{\smallskip} 
4202430102 & 59539.1 & <0.05 & 1.03$^{+0.02}_{-0.02} $ & 3.67$^{+0.09}_{-0.09}  $ & 1.22 & 351 \\\noalign{\smallskip} 
4202430103 & 59540.5 & <0.018 & 1.12$^{+0.03}_{-0.02} $ & 3.19$^{+0.08}_{-0.08}  $ & 1.11 & 379 \\\noalign{\smallskip} 
4202430104 & 59541.5 & <0.007 & 1.06$^{+0.02}_{-0.02} $ & 3.29$^{+0.06}_{-0.05}  $ & 1.24 & 397 \\\noalign{\smallskip}
4202430105 & 59542.5 & <0.4 & 1.14$^{+0.07}_{-0.05} $ & 2.40$^{+0.12}_{-0.13}  $ & 1.16 & 154 \\\noalign{\smallskip}
4202430106 & 59543.6 & <0.1 & 1.06$^{+0.02}_{-0.02} $ & 2.66$^{+0.05}_{-0.07}  $ & 1.08 & 328 \\\noalign{\smallskip}
4202430107 & 59544.4 & <0.015 & 1.12$^{+0.03}_{-0.02} $ & 2.61$^{+0.05}_{-0.07}  $ & 0.95 & 364 \\\noalign{\smallskip}
4202430108 & 59545.5 & <0.03 & 1.02$^{+0.08}_{-0.08} $ & 2.77$^{+0.18}_{-0.14}  $ & 0.86 & 42 \\\noalign{\smallskip}
4202430109 & 59550.7 & <0.006 & 0.87$^{+0.02}_{-0.02} $ & 2.19$^{+0.05}_{-0.05}  $ & 1.18 & 404 \\\noalign{\smallskip}
4202430110 & 59551.7 & <0.007 & 0.69$^{+0.04}_{-0.04} $ & 2.12$^{+0.10}_{-0.10}  $ & 1.14 & 269 \\\noalign{\smallskip}
4202430111 & 59552.5 & <0.003 & 0.88$^{+0.02}_{-0.02} $ & 2.03$^{+0.06}_{-0.06}  $ & 1.19 & 355 \\\noalign{\smallskip}
4202430112 & 59553.1 & <0.11 & 0.91$^{+0.12}_{-0.10} $ & 1.28$^{+0.13}_{-0.13}  $ & 0.90 & 46 \\\noalign{\smallskip}
4202430113 & 59555.5 & <0.11  & 0.92$^{+0.10}_{-0.09} $ & 1.12$^{+0.09}_{-0.09}  $ & 1.03 & 103 \\\noalign{\smallskip}
4202430114 & 59556.7 & <0.04 & 0.69$^{+0.07}_{-0.07} $ & 1.42$^{+0.09}_{-0.13}  $ & 0.97 & 96 \\\noalign{\smallskip}
4202430116 & 59559.5 & <0.07 & 0.99$^{+0.08}_{-0.07} $ & 2.24$^{+0.14}_{-0.14}  $ & 1.03 & 157 \\\noalign{\smallskip}
4202430117 & 59560.8 & <0.06 &0.92$^{+0.10}_{-0.07} $ & 1.87$^{+0.14}_{-0.16}  $ & 0.87 & 115 \\\noalign{\smallskip}
4202430118 & 59561.6 & <0.10 & 0.90$^{+0.09}_{-0.09} $ & 1.45$^{+0.12}_{-0.11}  $ & 1.09 & 116 \\\noalign{\smallskip}
4202430119 & 59563.3 & <0.008 & 0.97$^{+0.04}_{-0.04} $ & 1.98$^{+0.10}_{-0.09}  $ & 1.11 & 171 \\\noalign{\smallskip}
4202430121 & 59571.5 & <0.13  & 0.74$^{+0.17}_{-0.16} $ & 2.14$^{+0.4 }_{-0.3 }  $ & 1.41 & 15 \\\noalign{\smallskip}
4202430141 & 59602.6 & <0.07 & 1.09$^{+0.11}_{-0.08} $ & 2.15$^{+0.18}_{-0.18}$ & 1.05 & 50 \\ \hline\noalign{\smallskip}
\end{tabular}
% }
\tnote{(a)} Column density intrinsic to the SMC and the source, most values are not well constrained and are consistent with an upper limit. For the fit the Galactic column density was fixed to a value of 4\hcm{20}.
\end{threeparttable}
\end{table*}

\begin{figure*}
\begin{center}
%\resizebox{\hsize}{!}{
  \includegraphics[width=\textwidth,clip=]{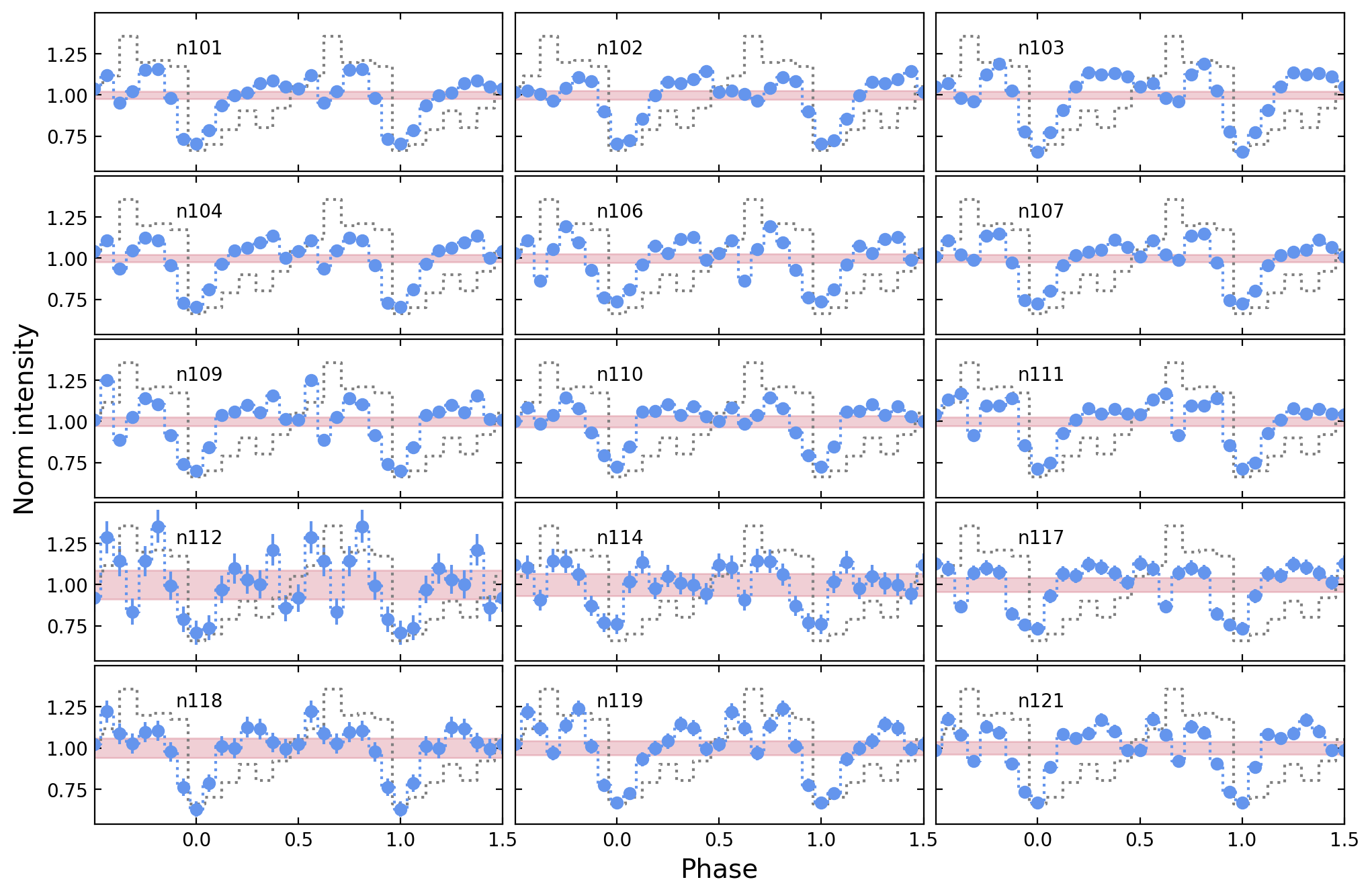}
%  }
\end{center}
  \caption{Pulse profiles (0.5--8.0 keV) of all \nicer observations where significant pulsations were detectable. Labels within each panel denote the last three digits nicer obsid number, i.e. 4202430XXX.
  All profiles are normalised to the average intensity and the minimum is shifted to zero phase. Shaded horizontal regions denote the limit of statistical significant variability above the constant level hypothesis. For comparison the 2016 pulse profile as measured by \cxo is plotted with grey dotted line in each panel.}
  \label{fig:PPs}
\end{figure*}

\begin{figure*}
 \includegraphics[width=\textwidth]{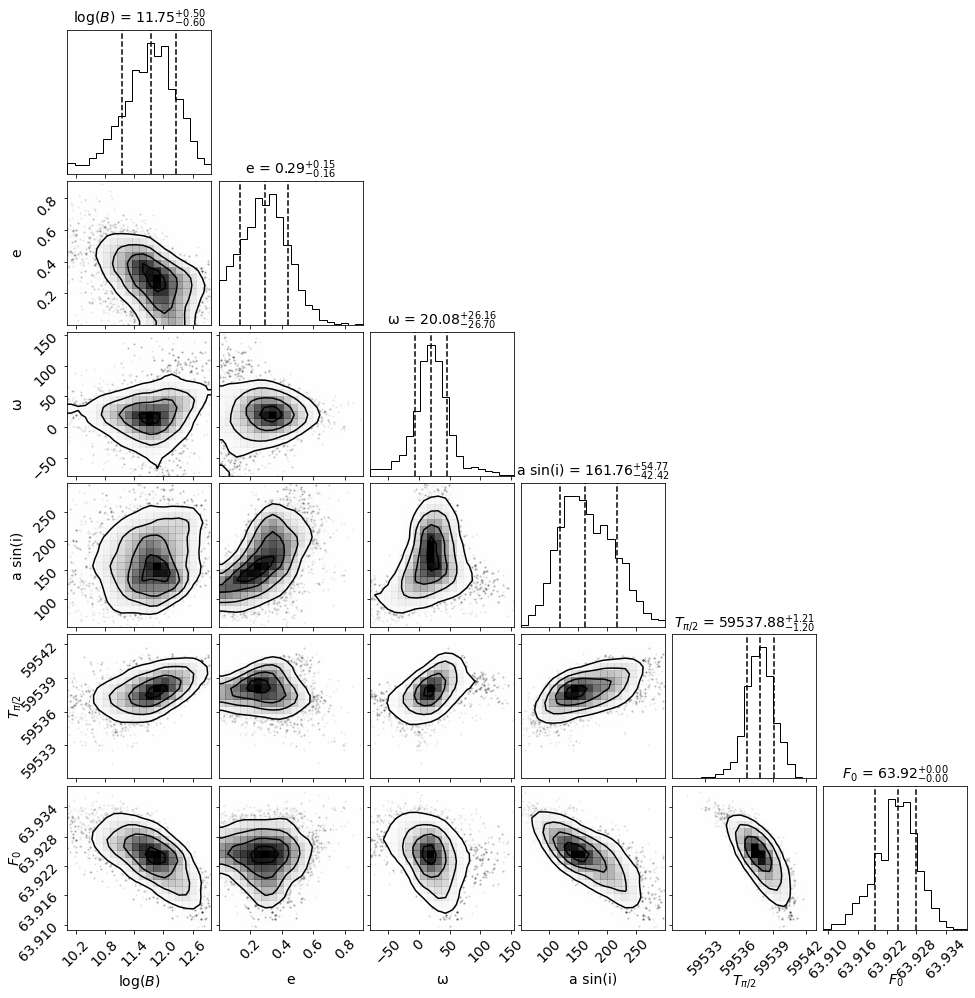}
    \caption{Corner plot for \sxp using the model described in the text. We plot the logarithm of the polar magnetic field strength $B$ in G. The eccentricity ($e$), the longitude of periastron in degrees ($\omega$) and the projected semi-major axis in light-sec ($a \, \sin i$), the time of a mean longitude of 90 degrees $T_{\rm \pi/2}$ and the pulsar frequency $F_0$ in mHz. The reference epoch for $F_0$ is the start of the \nicer monitoring, i.e. MJD 59538.5.}
    \label{fig:corner}
\end{figure*}

\end{appendix}

\end{document}